\documentclass[%
 reprint,
 amsmath,amssymb,
 aps,
 prl,
 floatfix,
]{revtex4-2}

\usepackage{graphicx}
\usepackage{dcolumn}
\usepackage{bm}
\usepackage{siunitx}
\usepackage[version=4]{mhchem}

\begin{document}

\preprint{SingleContact/0.0}

\title{Single-Collision Statistics Reveal a Global Mechanism Driven by Sample History for Contact Electrification in Granular Media}

\author{Galien Grosjean}
 \email{galienmariep.grosjean@ist.ac.at}
\author{Scott Waitukaitis}%
\affiliation{%
Institute of Science and Technology Austria (ISTA), Lab Building West, Am Campus 1, 3400 Klosterneuburg, Austria
}%

\date{\today}

\begin{abstract}

Models for same-material contact electrification in granular media often rely on a local charge-driving parameter whose spatial variations lead to a stochastic origin for charge exchange. Measuring the charge transfer from individual granular spheres after contacts with substrates of the same material, we find instead a `global' charging behavior, coherent over the sample's whole surface. Cleaning and baking samples fully resets charging magnitude and direction, which indicates the underlying global parameter is not intrinsic to the material, but acquired from its history.  Charging behavior is randomly and irreversibly affected by changes in relative humidity, hinting at a mechanism where adsorbates, in particular water, are fundamental to the charge-transfer process.

\end{abstract}

\maketitle

Contact electrification (CE), the transfer of electrical charge when objects touch, plays a crucial role in granular media \cite{lacks2019}. In nature, ice crystals in thunder clouds or ash particles in volcanic plumes collide and charge to help create spectacular displays of lightning \cite{Gilbert.1991v, Pahtz.2010}.  In industrial settings, e.g.~fluidized beds \cite{Grosshans.2016} or pharmaceutical plants \cite{rescaglio2019}, CE adversely affects adhesion and flow, but can also be harnessed for filtration \cite{Liu.2020}.  In grain silos, sparks from charged grains can ignite deadly explosions \cite{Abbasi.2007}. Charged dust is important for space exploration, as landers and rovers must be engineered to withstand its accumulation \cite{Calle.2011}. Further away still, charged grains are suspected to play an essential role in rocky planet formation, speeding up the process sufficiently to allow Earth-like planets to exist \cite{Steinpilz.2020, Singh.2018, Lee.2015}.

Regarding what causes CE, in particular for insulators where the effect is strongest, there is no consensus on the mechanism or the species transferred \cite{lacks2019}. With different materials, it is widely assumed that charge transfer is driven by a material parameter \cite{Matsuyama.2006, Ireland.2010, Grosshans.2017}. This model is `global' in that charge-transfer does not vary with the location of the contact. In granular media, charging occurs between grains of the same material, seemingly precluding a global mechanism. Hypotheses to overcome this historically resort to a `local' picture for charge exchange, \textit{i.e.}~where the charge-driving parameter varies over the surface \cite{Lowell.1986b,  Lacks.2008, Kok.2009, Forward.2009bs}. This parameter would average out over large scales to render grains identical globally, but nonetheless change sufficiently over the scale of contacts to permit transfer. Prominent recently are `patch models', where surfaces are thought to consist of nanoscale donor/acceptor regions and charging arises stochastically from exchange between these \cite{apodaca2010, Yu.2017, xie2016, grosjean2020}. Relevant to any mechanism is the omnipresent influence of adsorbed surface water. For global models, water is generally seen as providing a conductive path that amplifies some other underlying charge-driving mechanisms \cite{Zheng.2014}. For local models, `islands' of adsorbed water have been implicated as the actors that define patches \cite{Zhang.2015, grosjean2020, lee2018, xie2016}.
 
While a local model might seem necessary to explain same-material CE in granular media, we are not aware of any experiments that directly demonstrate its occurrence. In principle, all that is needed are samples that are as identical as possible, a careful preparation protocol to keep them so, and a statistically significant number of charge-exchange measurements at random surface locations. Local models predict this should lead to charge-exchange distributions with zero average, while for any global mechanism it would be non-zero. Yet this is not an easy task. It is straightforward to probe bulk granular CE with Faraday cups, but this yields no information on individual grains \cite{LaMarche.2009}. Some experiments address single grains, but are not precise enough to measure charge exchange \cite{Waitukaitis.2013un5, Waitukaitis.2014, Steinpilz.2020, Carter.2020}. A handful measure charge exchange, but with different materials \cite{Matsuyama.1995, Matsuyama.2003, Watanabe.2006, Haeberle.2018} or with centimeter-scale objects \cite{Xie.2013, xie2016, Hu.2012, Zhang.2015} to enhance magnitude. We are not aware of any experiments with sufficient resolution and flexibility to gather comprehensive statistics of same-material CE at the scale of a single grain.   

\nocite{kline2020}

\begin{figure*}[ht!]
    \includegraphics[width=17.8cm]{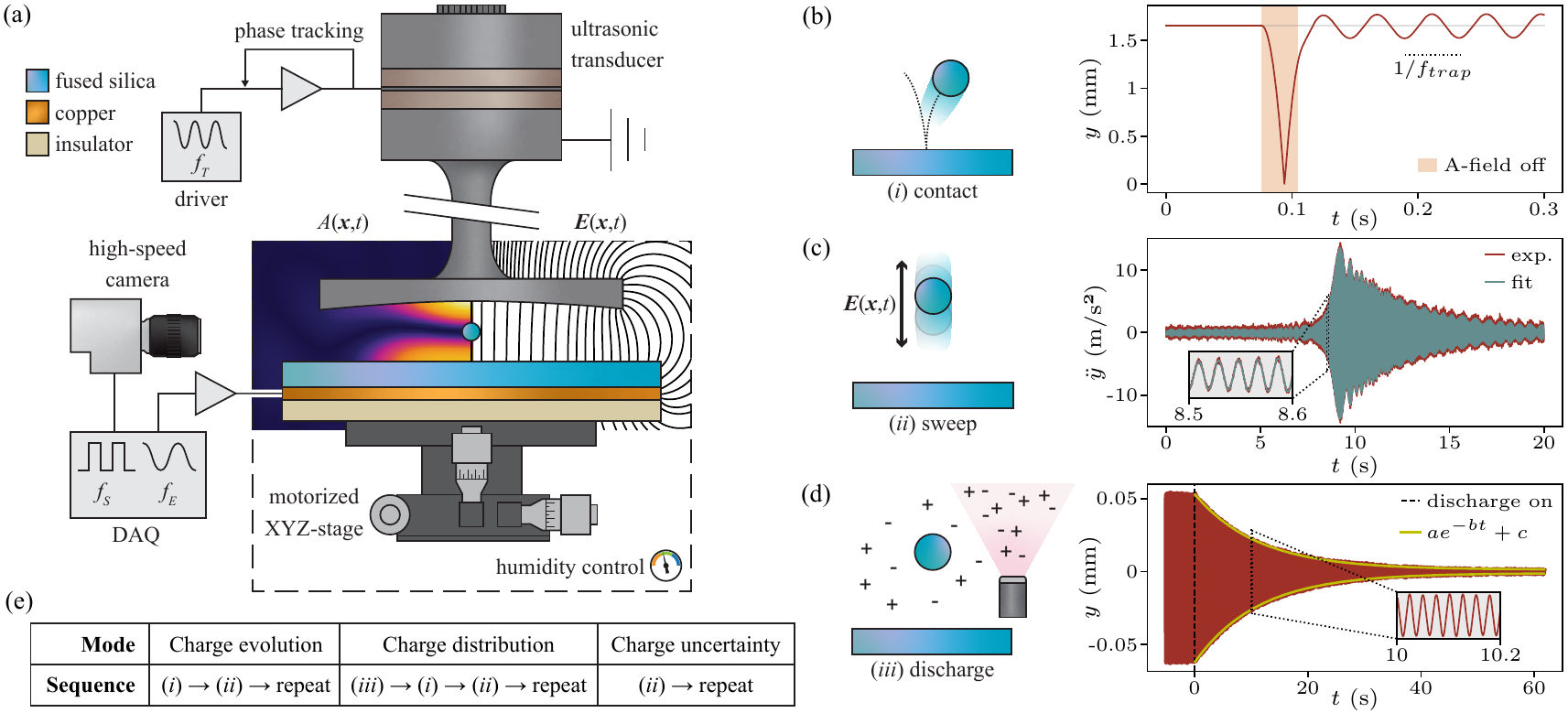}
    \caption{\label{fig1} \textbf{Setup and protocol.} (a) The setup consists of a Langevin transducer above the target substrate and electrode. The spherical particle levitates in the node of the acoustic standing wave. (b) Charge-exchanging contacts are initiated by briefly interrupting the acoustic field, causing the sphere to bounce exactly once on the surface before we `catch' it.  (c) To measure charge, we frequency sweep a spatially uniform applied electric field through the sphere's resonance and track its position with a high-speed camera.  Fitting the acceleration to Eq.~(\ref{2ndlaw}) yields the charge. (d) Trajectory of a charged sphere in response to a harmonic E-field with the discharger off ($t<0$), and then on ($t>0$); fitting for $t>0$ to an exponential gives a time constant of $\sim$8 s.  
    (e) The three tasks shown in (b)-(d), contact (\textit{i}), sweep (\textit{ii}) and discharge (\textit{iii}), can be combined in different ways depending on the measurement mode. For further details on the setup and videos demonstrating contacts, charge measurement, and discharge, see the Supplemental Material \cite{SupplMat}.
    }
\end{figure*}
 
In this Letter, we dissect the global \textit{vs.}~local nature of same-material granular CE by further pioneering charge measurement via acoustic levitation \cite{andrade2018, lee2018, kline2020, Harvey.2022}, which enables exquisite charge resolution and automated contacts without physical handling. Observing the charge evolution over sequential contacts and the charge-exchange distributions of initial contacts, we demonstrate that the symmetry-breaking parameter is global. We find that this parameter is not inherent to individual samples, but acquired during their history---merely recleaning and rebaking samples can flip the charging direction. Considering the ubiquitous influence of adsorbed water, we vary relative humidity (RH), expecting to uniformly affect charging. Instead, we find random shifts to the exchange: the same change in RH can cause charging to either increase or decrease, and irreversibly so. Our results suggest that same-material granular CE is determined by and extremely sensitive to environmental history, pointing to adsorbates---and in particular water---as the charge-driving agents.

Our samples are research-grade fused silica (\ce{SiO_2}) spheres and substrates, carefully selected to be as pure and identical as possible.  Both are made from a single traceable source material, Heraeus Spectrosil\textsuperscript{\textregistered} 2000, which limits bulk impurities to parts per billion.  The spheres (Sandoz Fils SA., grade 25) have diameters $D=500\pm\SI{1}{\um}$. The substrates (UQG Optics Ltd WFS-252) are disks with $\SI{25}{mm}$ diameter and $\SI{6}{mm}$ thickness. AFM topography measurements on spheres/substrates reveal roughness on the order of $\SI{4}{nm}$ and $\SI{1}{nm}$, respectively. Spheres and substrates are subjected to a rigorous cleaning protocol before experiments: first sonicating for 30 minutes each in acetone ($>\,$99.5\,\%), methanol ($>\,$99.9\,\%), and Milli-Q\textsuperscript{\textregistered} water, and then baking overnight at 300$^{\circ}$C. A particular sphere/substrate pair always undergoes this protocol jointly, \textit{i.e.}~together in the same beakers with the same solvent at each step.  Immediately after baking, samples enter a temperature ($\pm$$2^{\circ}$C) and RH ($\pm$1\%) regulated environment, also fed by Milli-Q\textsuperscript{\textregistered} water.

The experimental apparatus is illustrated in Fig.~\ref{fig1}(a), and builds upon the acoustic levitation technique introduced in Refs.~\cite{lee2018, kline2020}. We levitate a sphere using an ultrasonic standing wave created by a resonant Langevin transducer suspended above our target substrate. To initiate a contact, we briefly interrupt the acoustic field, with the duration ($\sim$\,$\SI{25}{\milli\second}$) tuned so that the sphere falls and bounces exactly once before it is recaught in the trap; see Fig.~\ref{fig1}(b) and Supplemental Material, Video 1 \cite{SupplMat}.  To measure charge, a spatially uniform electric field is AC-swept to pass through the natural frequency of the sphere in the acoustic trap ($f_{\mathrm trap} \approx\SI{50}{\hertz}$). We record the sphere's motion with a high-speed camera (Phantom VEO 640L) and use particle tracking to obtain its vertical position as a function of time, $y(t)$. Newton's second law projected on the vertical direction can be written
\begin{equation}
\label{2ndlaw}
    \ddot{y}
    = -g - a \sin 2ky - 2 \beta_0 \dot{y} - 2 \beta_1 |\dot{y}| \dot{y} + Q E(t) / m.
\end{equation}
The electric field, $E(t)$, the acoustic wavenumber, $k$, and the sphere mass, $m$, are known, and the first and second derivatives of $y$ can be numerically calculated. The unknowns are the acoustic amplitude, $a$, the linear and quadratic damping coefficients, $\beta_0$ and $\beta_1$, arising from air drag, and charge, $Q$, which we obtain by fitting. Typical acceleration data and a fit are shown in Fig.~\ref{fig1}(c) (see also Supplemental Material, Video 1 \cite{SupplMat}).

Several advances beyond previous works \cite{Lee.2015, kline2020} are required for our purposes.  First, we must be able to change the location of contact on both the sphere and the substrate. With the sphere, symmetry prevents any preferred orientation, causing it to rotate while levitating such that contacts occur at a random locations. This rotation is visible in Supplemental Material, Video 2, and in the Supplemental Material we estimate the frequency to be of the order of 100~Hz \cite{SupplMat}. For the substrate, we incorporate a piezo-driven XYZ-stage to laterally displace it between contacts [Fig.~\ref{fig1}(a)]. We move it in a square spiral with steps of $\SI{20}{\um}$, just larger than the estimated contact diameter ($d\approx \SI{19.7}{\um}$). Second, to carry out experiments with the same initial (zero charge) conditions, we introduce a discharge mechanism. We place a photo\-ionizer (Hamamatsu L12645) in the chamber directed away from the sphere/substrate, which enhances the conductivity of the surrounding air to cause rapid discharge. Figure~\ref{fig1}(d) shows how the steady trajectory of a sphere shaken harmonically at constant amplitude quickly decays after the device is turned on.  Fitting to an exponential yields a time constant of $\sim$$\SI{8}{\second}$ (see Supplemental Material, Video 1 \cite{SupplMat}).

\begin{figure}[t]
    \includegraphics[width=8.6cm]{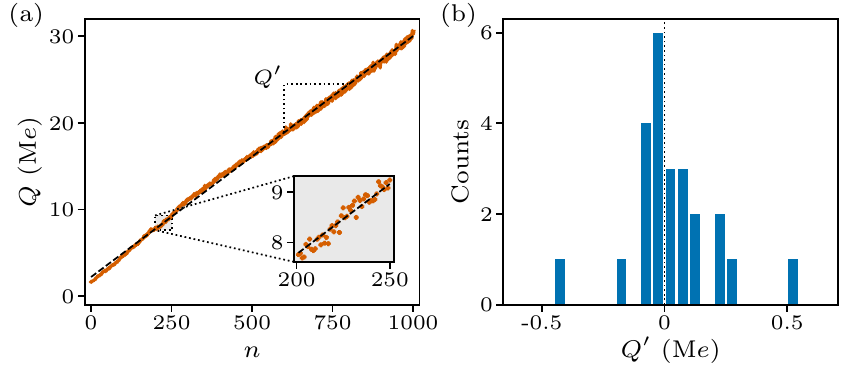}
    \caption{\label{fig2} \textbf{Charge evolution.} (a) The charge, $Q$, of a sphere is measured after $n$ contacts with the substrate and with no discharge between, \textit{i.e.}~in `charge evolution mode,' for a total of 1000 bounces. The steady charging rate, $Q'$, indicates a global difference between this sphere/substrate pair.  (b) Distribution of $Q'$ measured for 25 sphere/subtrate pairs, which is centered on zero; this indicates that there is no systematic difference between spheres and substrates.}
\end{figure}

The capacity to perform (\textit{i}) contact, (\textit{ii}) charge measurement, and (\textit{iii}) discharge gives us access to otherwise unattainable modes of experimentation. The most direct is `charge evolution mode', e.g.~in Fig.~\ref{fig2}(a), where we perform 1000 cycles of contact then charge measurement (\textit{i}$\,\rightarrow\,$\textit{ii}$\,\rightarrow\,$repeat). As is clear, the sphere's charge in this instance marches steadily upward at a constant rate per collision, $dQ/dn\equiv Q'$. In the standard patch model, net charge is exchanged in a single collision due to fluctuations in the number of charge donor/acceptor pairs facing each other at the contact location, but over many locations the average is predicted to approach zero~\cite{apodaca2010, grosjean2020}. Hence, the data in Fig.~\ref{fig2}(a) already indicate a global mechanism driving exchange in this sphere/substrate pair.  Similar trends were seen before, but this implication was missed  \cite{lee2018}. If all spheres charged with the same sign against all substrates, one could argue that they differ in an intrinsic way, but Fig.~\ref{fig2}(b) shows this is not the case.  Calculating the distribution of the rates $Q'$ for an ensemble of sphere/substrate pairs shows that they are centered around zero---each sphere is globally different from each substrate, but the average difference is zero.

\begin{figure}[t]
    \includegraphics[width=8.6cm]{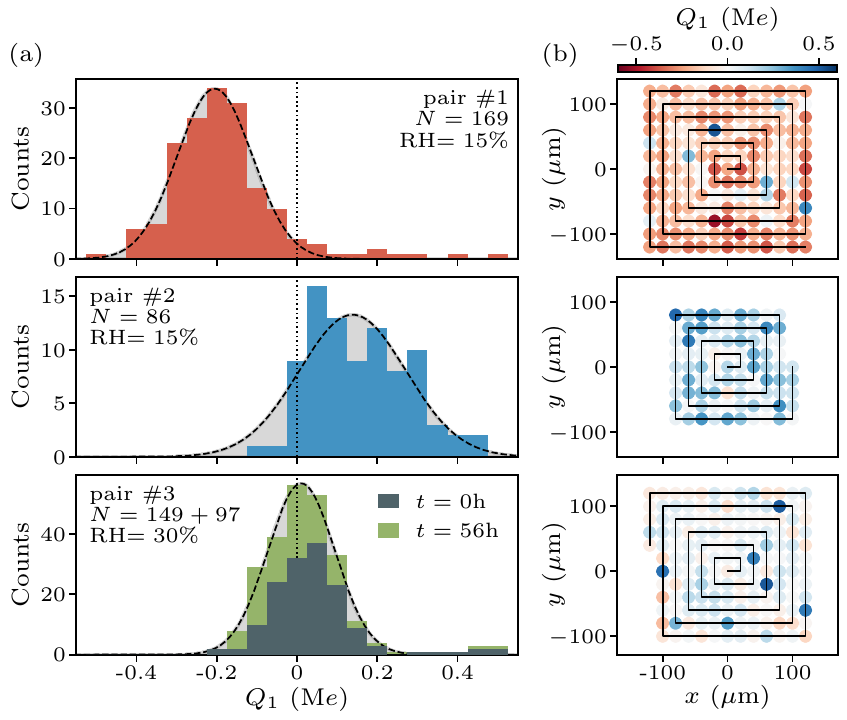}
    \caption{\label{fig3} \textbf{Charge distributions.} (a) In `charge distribuiton mode,' the sphere charge after the first contact, $Q_1$, is repeatedly measured for a given sphere/substrate pair by discharging the system between collisions. The median of the distribution can be either negative (pair \#1), positive (pair \#2) or close to zero (pair \#3). With pair \#3, the distribution is composed of two sets of measurements, taken 56 hours apart (colored dark gray and light green), showing no significant drift of the distribution with time.  Gaussian fits are shown only as a guide to the eye. (b) Between contacts, the substrate is moved along a square spiral. Plotting $Q_1$ \textit{vs.}~the contact location on the substrate, no clear trend can be identified with either space or time.}
\end{figure}

Beyond charge evolution, we can also measure the distribution of charge exchange, $Q_1$, for the initial (fully discharged) contact of a sphere/substrate pair. In this `charge distribution mode', we cycle over discharging, performing a contact, and then measuring charge (\textit{iii}$\,\rightarrow\,$\textit{i}$\,\rightarrow\,$\textit{ii}$\,\rightarrow\,$repeat).  As Fig.~\ref{fig2} shows, the typical magnitude of charge exchange is $\sim$$\,10^5\,e$, and as we explain regarding `charge uncertainty mode' in the Supplemental Material \cite{SupplMat} our measurement uncertainty is $\lesssim$$\,10^3\,e$---two orders of magnitude lower. With this level of resolution, charge-exchange distributions, even with our small samples and identical materials, are easily resolved. Typical results for three sphere/substrate pairs are shown in Fig.~\ref{fig3}(a). As is immediately observed, the distributions are not constrained to be centered around zero, as a local model would require. The distribution for sample pair \#1 is predominantly negative,  pair \#2 positive, and pair \#3 close to zero. Though we do not delve into their shapes \cite{Haeberle.2018}, distributions are often approximately Gaussian, with widths of around $10^5\,e$.  We confirm that distributions are stable over time by repeating the same measurement several days apart. For instance, the distribution of pair \#3 is comprised of two sets of measurements, taken 56 hours apart and shown in different colors. Neither the median value nor the standard deviation display any discernible change.

What we learned from charge evolution [Fig.~\ref{fig2}(a)] is thus reconfirmed by the charge distributions: the charging between a particular sphere/substrate pair is driven by a global, not local, parameter.  To make this point even stronger, Fig.~\ref{fig3}(b) shows $Q_1$ as a function of the contact location on the substrate, following the square spiral from the center outward. Here the global charging behavior becomes visually apparent---a substrate that charges positive/negative does so over large regions of its surface. Positive/negative regions are not spatially correlated, and no drift over time occurs. The fact that charge distributions can be strictly positive or negative also suggests that the first contact does not play any special role in breaking symmetry. Electric fields could certainly influence charge exchange~\cite{lee2018, Pahtz.2010}, however charging direction is preserved after discharge, suggesting that polarization from the particle's own field is not what determines the sign during subsequent collisions~\cite{yoshimatsu2017}.

\begin{figure}[t]
    \includegraphics[width=8.6cm]{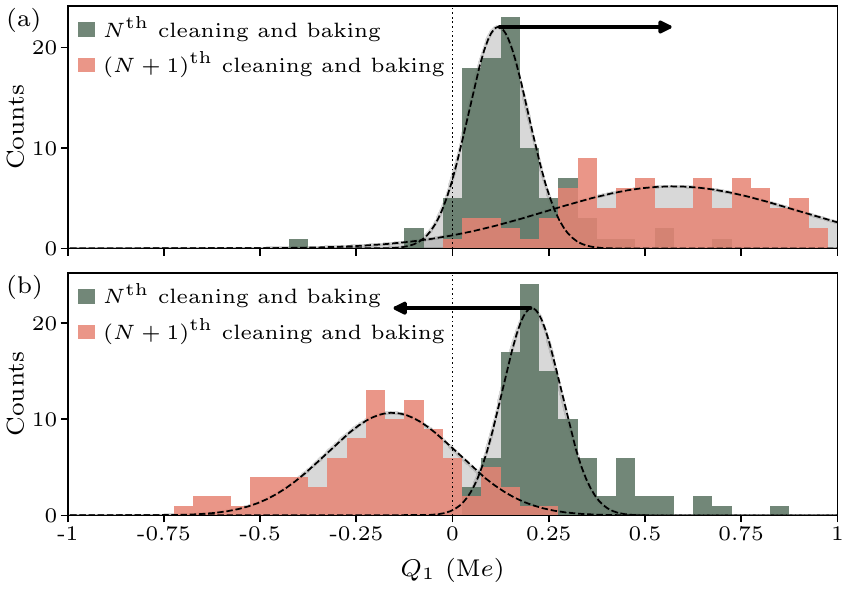}
    \caption{\label{fig4} \textbf{Resetting charging via recleaning and rebaking.} If we reclean and rebake a sphere/substrate pair, their charging behavior is `reset.' Every time, the median and width of the distribution is randomly altered; even the sign can be flipped.  This indicates that the global charge-driving parameter is acquired, not intrinsic.}
\end{figure}

We now perform experiments to uncover the nature of the charge-driving parameter, starting with the question: is it intrinsic to a given sphere/substrate pair, or acquired during their history? To answer this, we measure a $Q_1$ distribution for a particular pair, and then reclean and rebake them together and measure the distribution again. As can be seen in Fig.~\ref{fig4}, repreparing samples changes the median value and width of the original distribution. The difference is such that it would be impossible to tell whether the same pair has been used or a different one---the conditions are entirely reset. Most notably, as in Fig.~\ref{fig4}(b), the sign of charging can be flipped. We conclude that the global parameter driving charging is an acquired one, with the most likely candidate being surface adsorbates. Considering that baking at a few hundred degrees removes most (though not all \cite{Zhuravlev.2000}) organic adsorbates, the implicated species are likely acquired afterward when samples enter the experimental chamber, where controlled RH is maintained. We thus consider the possibility, as have many others recently, that adsorbed surface water is driving the charging \cite{Lee.2015, Burgo.2016, Gil.2019, Jungmann.2022x, Zhang.2015, xie2016}.

\begin{figure}[t]
    \includegraphics[width=8.6cm]{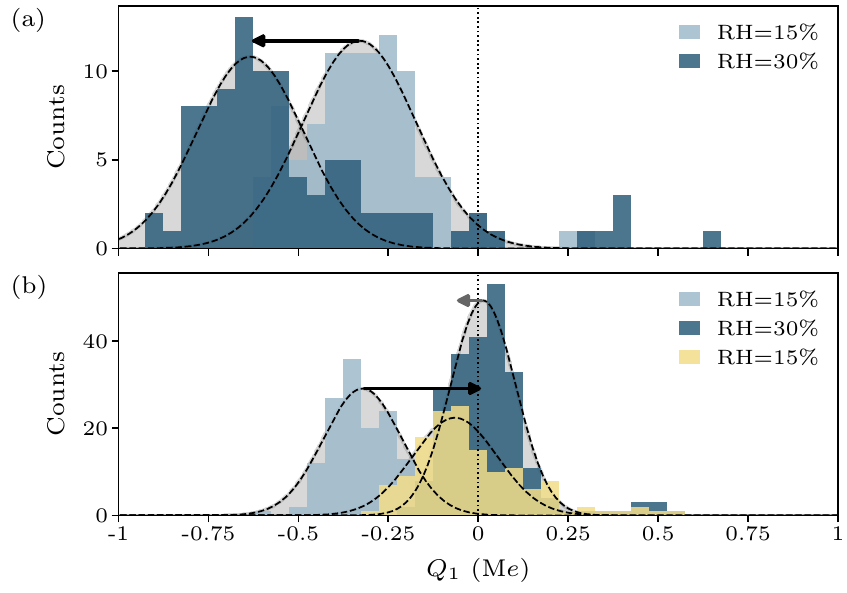}
    \caption{\label{fig5} \textbf{Charge shifts via humidity changes.} Changing the RH of the experimental chamber (from 15\,\% to 30\,\%)
    while a sphere/substrate pair are present causes their charging distributions to shift.  
    This shift can either (a) increase or (b) decrease the charging magnitude. Such shifts are not reversible, as shown in (b) where the RH was decreased back to 15\,\%.  These effects are similar to those observed after recleaning and rebaking, though no clear sign flips were observed. 
    }
\end{figure}

To find out how, we measure $Q_1$ distributions for sample pairs before/after they jointly experience changes of RH. Several results in the literature, using both different and same materials, have indicated that ensemble averages for charge-exchange magnitude reach a maximum at $\sim$\,30\,\% RH~\cite{rescaglio2019,xie2016,schella2017}. One might expect, then, that varying RH would uniformly affect CE for our sphere/substrate pairs, increasing the magnitude when moving closer to the optimum and decreasing it when moving away. This is not what is observed. Increasing RH from 15 to 30\,\% causes random shifts to charging magnitude---about half of the time it is increased [Fig.~\ref{fig5}(a)], and half of the time it is decreased [Fig.~\ref{fig5}(b)]. The shift can be large---often comparable in magnitude to the distribution widths. As shown in Fig.~\ref{fig5}(b), lowering the RH back to 15\,\% does not undo the shift---the changes are not reversible.  Shifts are strongest between 15\,\% to 30\,\% RH, and increasing RH beyond 30\,\% has relatively little effect.  We observe that exposure to RH $\gg\,$30\,\% largely diminishes any shifts thereafter.

If adsorbed water were merely a conduit for some other underlying global mechanism, then the addition/removal via RH should affect charging uniformly. Instead, we observe a significant, random and irreversible alteration of the charging behavior. The shifts observed due to RH are typically smaller than those caused by recleaning and rebaking, however they bear qualitative similarities. While no statistically significant sign flips were observed due to RH, charging sometimes went from clearly positive/ negative to essentially zero [Fig.~\ref{fig5}(b)], or \textit{vice versa}. If adsorbed water drives the charging behavior, then we indeed expect the changes from RH to be weaker than those from recleaning and especially rebaking, after which samples would be presumably devoid of most water  \cite{Zhuravlev.2000} and hence completely `reset', consistent with what we see.

On the other hand, our data also show that water is not merely an actor in a local patch model.  Some new type of `global patch model' could explain our data, provided that water-patch coverage can (1) be different from one sample to the next at a fixed RH (to explain Figs.~\ref{fig2} \& \ref{fig3}) and (2) evolve differently for each sample with preparation and/or RH (to explain Figs.~\ref{fig4} \& \ref{fig5}).  However, if this is the correct interpretation, it requires a new aspect of surface water to be considered---adsorption hysteresis. While we found no discussion of this phenomenon in the CE literature, it is well documented in other contexts \cite{Sullivan.2007cat, Taqvi.1999, Rudisill.1992, Barnette.2012, Qi.2005abc}. It seems to be particularly important when multiple adsorbates compete on a surface. For example, when water coadsorbs on an \ce{SiO_2} surface with different alcohols, the water surface coverage need not have a single value for a given RH, and evolves differently with RH depending on history \cite{Barnette.2012}. The magnitude of the effect can be such that, under identical environmental conditions, two same-material surfaces with different histories can have surface water coverages that differ by up to a monolayer. Significant adsorption hysteresis is also known to occur in porous materials, including porous glasses, gels and polymers~\cite{kierlik2001,kierlik2002,chen2018}. Considering that the scale of charge exchange in CE typically only requires about one in every $10^4$ surface atoms/molecules ($\sim$10$^{-4}$ monolayers) to participate, and that water in atmospheric conditions coadsorbs with a complex mixture of many other molecules, effects from adsorption hysteresis cannot by any means be excluded. Nonetheless, to clearly establish that water drives charge exchange would require to correlate charging behavior with direct measurements of adsorbed water. Such measurements would have to be precise enough to resolve minute differences, potentially down to the submonolayer scale.

Having recently published theoretical work on same-material CE based on a local, patch-driven framework \cite{grosjean2020}, we embarked upon these experiments with the expectation that signatures of a local model could be observed. However, despite extreme care with regard to sample purity and preparation, we only find evidence of a global mechanism---the tendency to charge positive/negative does not average to zero from one contact location to the next, but is stable over large length scales. Our data tells us the charge-driving parameter is acquired during sample history; it can be reset by cleaning and baking, and randomly and irreversibly shifted via RH. These observations are difficult to reconcile with mechanisms based on intrinsic parameters, including: work functions, dielectric constants, specific heats, Seebeck coefficients, surface roughness, flexoelectric constants, piezoelectric constants, mechanochemistry, \textit{etc.} Polarization~\cite{Pahtz.2010, yoshimatsu2017} does not seem to cause the initial symmetry breaking, though we cannot rule out this effect without additional studies of contacts under applied electric fields. The most consistent mechanism we can propose is that the global charge-driving parameter is related to adsorbates acquired during a sample's history, in particular water. Though many other authors have proposed that water plays an important role, our data suggest a new twist---namely that minute deviations in conditions during water adsorption lead to global differences in water coverage, which drive charging. Such a twist is not outside reason considering the unpredictability and irreproducibility of CE generally \cite{lacks2019, Lacks.2012}. Even more so when one considers the well-documented existence of adsorption hysteresis causing coverage differences up to a full monolayer \cite{Sullivan.2007cat, Taqvi.1999, Rudisill.1992, Barnette.2012, Qi.2005abc}. Further investigations that attempt to correlate surface water coverage and CE directly would be extremely valuable in testing this hypothesis.

We would like to thank Troy Shinbrot, Victor Lee and Daniele Foresti for helpful discussions. This project has received funding from the European Research Council Grant Agreement No. 949120 and from the the Marie Sk\l{}odowska-Curie Grant Agreement No. 754411 under the European Union's Horizon 2020 research and innovation program.

\bibliographystyle{apsrev4-2}
\bibliography{apssamp}

\begin{thebibliography}{54}%
\makeatletter
\providecommand \@ifxundefined [1]{%
 \@ifx{#1\undefined}
}%
\providecommand \@ifnum [1]{%
 \ifnum #1\expandafter \@firstoftwo
 \else \expandafter \@secondoftwo
 \fi
}%
\providecommand \@ifx [1]{%
 \ifx #1\expandafter \@firstoftwo
 \else \expandafter \@secondoftwo
 \fi
}%
\providecommand \natexlab [1]{#1}%
\providecommand \enquote  [1]{``#1''}%
\providecommand \bibnamefont  [1]{#1}%
\providecommand \bibfnamefont [1]{#1}%
\providecommand \citenamefont [1]{#1}%
\providecommand \href@noop [0]{\@secondoftwo}%
\providecommand \href [0]{\begingroup \@sanitize@url \@href}%
\providecommand \@href[1]{\@@startlink{#1}\@@href}%
\providecommand \@@href[1]{\endgroup#1\@@endlink}%
\providecommand \@sanitize@url [0]{\catcode `\\12\catcode `\$12\catcode
  `\&12\catcode `\#12\catcode `\^12\catcode `\_12\catcode `\%12\relax}%
\providecommand \@@startlink[1]{}%
\providecommand \@@endlink[0]{}%
\providecommand \url  [0]{\begingroup\@sanitize@url \@url }%
\providecommand \@url [1]{\endgroup\@href {#1}{\urlprefix }}%
\providecommand \urlprefix  [0]{URL }%
\providecommand \Eprint [0]{\href }%
\providecommand \doibase [0]{https://doi.org/}%
\providecommand \selectlanguage [0]{\@gobble}%
\providecommand \bibinfo  [0]{\@secondoftwo}%
\providecommand \bibfield  [0]{\@secondoftwo}%
\providecommand \translation [1]{[#1]}%
\providecommand \BibitemOpen [0]{}%
\providecommand \bibitemStop [0]{}%
\providecommand \bibitemNoStop [0]{.\EOS\space}%
\providecommand \EOS [0]{\spacefactor3000\relax}%
\providecommand \BibitemShut  [1]{\csname bibitem#1\endcsname}%
\let\auto@bib@innerbib\@empty
\bibitem [{\citenamefont {Lacks}\ and\ \citenamefont
  {Shinbrot}(2019)}]{lacks2019}%
  \BibitemOpen
  \bibfield  {author} {\bibinfo {author} {\bibfnamefont {D.~J.}\ \bibnamefont
  {Lacks}}\ and\ \bibinfo {author} {\bibfnamefont {T.}~\bibnamefont
  {Shinbrot}},\ }\href {https://doi.org/10.1038/s41570-019-0115-1} {\bibfield
  {journal} {\bibinfo  {journal} {Nat. Rev. Chem.}\ }\textbf {\bibinfo {volume}
  {3}},\ \bibinfo {pages} {465} (\bibinfo {year} {2019})}\BibitemShut {NoStop}%
\bibitem [{\citenamefont {Gilbert}\ \emph {et~al.}(1991)\citenamefont
  {Gilbert}, \citenamefont {Lane}, \citenamefont {Sparks},\ and\ \citenamefont
  {Koyaguchi}}]{Gilbert.1991v}%
  \BibitemOpen
  \bibfield  {author} {\bibinfo {author} {\bibfnamefont {J.~S.}\ \bibnamefont
  {Gilbert}}, \bibinfo {author} {\bibfnamefont {S.~J.}\ \bibnamefont {Lane}},
  \bibinfo {author} {\bibfnamefont {R.~S.~J.}\ \bibnamefont {Sparks}},\ and\
  \bibinfo {author} {\bibfnamefont {T.}~\bibnamefont {Koyaguchi}},\ }\href
  {https://doi.org/10.1038/349598a0} {\bibfield  {journal} {\bibinfo  {journal}
  {Nature (London)}\ }\textbf {\bibinfo {volume} {349}},\ \bibinfo {pages}
  {598} (\bibinfo {year} {1991})}\BibitemShut {NoStop}%
\bibitem [{\citenamefont {Pähtz}\ \emph {et~al.}(2010)\citenamefont {Pähtz},
  \citenamefont {Herrmann},\ and\ \citenamefont {Shinbrot}}]{Pahtz.2010}%
  \BibitemOpen
  \bibfield  {author} {\bibinfo {author} {\bibfnamefont {T.}~\bibnamefont
  {Pähtz}}, \bibinfo {author} {\bibfnamefont {H.~J.}\ \bibnamefont
  {Herrmann}},\ and\ \bibinfo {author} {\bibfnamefont {T.}~\bibnamefont
  {Shinbrot}},\ }\href {https://doi.org/10.1038/nphys1631} {\bibfield
  {journal} {\bibinfo  {journal} {Nat. Phys.}\ }\textbf {\bibinfo {volume}
  {6}},\ \bibinfo {pages} {364 } (\bibinfo {year} {2010})}\BibitemShut
  {NoStop}%
\bibitem [{\citenamefont {Grosshans}\ and\ \citenamefont
  {Papalexandris}(2016)}]{Grosshans.2016}%
  \BibitemOpen
  \bibfield  {author} {\bibinfo {author} {\bibfnamefont {H.}~\bibnamefont
  {Grosshans}}\ and\ \bibinfo {author} {\bibfnamefont {M.~V.}\ \bibnamefont
  {Papalexandris}},\ }\href {https://doi.org/10.1016/j.powtec.2016.07.031}
  {\bibfield  {journal} {\bibinfo  {journal} {Powder Technol.}\ }\textbf
  {\bibinfo {volume} {301}},\ \bibinfo {pages} {1008} (\bibinfo {year}
  {2016})}\BibitemShut {NoStop}%
\bibitem [{\citenamefont {Rescaglio}\ \emph {et~al.}(2019)\citenamefont
  {Rescaglio}, \citenamefont {De~Smet}, \citenamefont {Aerts},\ and\
  \citenamefont {Lumay}}]{rescaglio2019}%
  \BibitemOpen
  \bibfield  {author} {\bibinfo {author} {\bibfnamefont {A.}~\bibnamefont
  {Rescaglio}}, \bibinfo {author} {\bibfnamefont {F.}~\bibnamefont {De~Smet}},
  \bibinfo {author} {\bibfnamefont {L.}~\bibnamefont {Aerts}},\ and\ \bibinfo
  {author} {\bibfnamefont {G.}~\bibnamefont {Lumay}},\ }\href
  {https://doi.org/10.1080/02726351.2018.1533606} {\bibfield  {journal}
  {\bibinfo  {journal} {Part. Sci. Technol.}\ }\textbf {\bibinfo {volume}
  {37}},\ \bibinfo {pages} {1024} (\bibinfo {year} {2019})}\BibitemShut
  {NoStop}%
\bibitem [{\citenamefont {Liu}\ \emph {et~al.}(2020)\citenamefont {Liu},
  \citenamefont {Cao}, \citenamefont {Huang}, \citenamefont {Chen},
  \citenamefont {Chen},\ and\ \citenamefont {Lai}}]{Liu.2020}%
  \BibitemOpen
  \bibfield  {author} {\bibinfo {author} {\bibfnamefont {H.}~\bibnamefont
  {Liu}}, \bibinfo {author} {\bibfnamefont {C.}~\bibnamefont {Cao}}, \bibinfo
  {author} {\bibfnamefont {J.}~\bibnamefont {Huang}}, \bibinfo {author}
  {\bibfnamefont {Z.}~\bibnamefont {Chen}}, \bibinfo {author} {\bibfnamefont
  {G.}~\bibnamefont {Chen}},\ and\ \bibinfo {author} {\bibfnamefont
  {Y.}~\bibnamefont {Lai}},\ }\href {https://doi.org/10.1039/c9nr08851b}
  {\bibfield  {journal} {\bibinfo  {journal} {Nanoscale}\ }\textbf {\bibinfo
  {volume} {12}},\ \bibinfo {pages} {437} (\bibinfo {year} {2020})}\BibitemShut
  {NoStop}%
\bibitem [{\citenamefont {Abbasi}\ and\ \citenamefont
  {Abbasi}(2007)}]{Abbasi.2007}%
  \BibitemOpen
  \bibfield  {author} {\bibinfo {author} {\bibfnamefont {T.}~\bibnamefont
  {Abbasi}}\ and\ \bibinfo {author} {\bibfnamefont {S.~A.}\ \bibnamefont
  {Abbasi}},\ }\href {https://doi.org/10.1016/j.jhazmat.2006.11.007} {\bibfield
   {journal} {\bibinfo  {journal} {J. Hazard. Mater.}\ }\textbf {\bibinfo
  {volume} {140}},\ \bibinfo {pages} {7 } (\bibinfo {year} {2007})}\BibitemShut
  {NoStop}%
\bibitem [{\citenamefont {Calle}\ \emph {et~al.}(2011)\citenamefont {Calle},
  \citenamefont {Buhler}, \citenamefont {Johansen}, \citenamefont {Hogue},\
  and\ \citenamefont {Snyder}}]{Calle.2011}%
  \BibitemOpen
  \bibfield  {author} {\bibinfo {author} {\bibfnamefont {C.}~\bibnamefont
  {Calle}}, \bibinfo {author} {\bibfnamefont {C.}~\bibnamefont {Buhler}},
  \bibinfo {author} {\bibfnamefont {M.}~\bibnamefont {Johansen}}, \bibinfo
  {author} {\bibfnamefont {M.}~\bibnamefont {Hogue}},\ and\ \bibinfo {author}
  {\bibfnamefont {S.}~\bibnamefont {Snyder}},\ }\href
  {https://doi.org/10.1016/j.actaastro.2011.06.010} {\bibfield  {journal}
  {\bibinfo  {journal} {Acta Astronaut.}\ }\textbf {\bibinfo {volume} {69}},\
  \bibinfo {pages} {1082} (\bibinfo {year} {2011})}\BibitemShut {NoStop}%
\bibitem [{\citenamefont {Steinpilz}\ \emph {et~al.}(2020)\citenamefont
  {Steinpilz}, \citenamefont {Joeris}, \citenamefont {Jungmann}, \citenamefont
  {Wolf}, \citenamefont {Brendel}, \citenamefont {Teiser}, \citenamefont
  {Shinbrot},\ and\ \citenamefont {Wurm}}]{Steinpilz.2020}%
  \BibitemOpen
  \bibfield  {author} {\bibinfo {author} {\bibfnamefont {T.}~\bibnamefont
  {Steinpilz}}, \bibinfo {author} {\bibfnamefont {K.}~\bibnamefont {Joeris}},
  \bibinfo {author} {\bibfnamefont {F.}~\bibnamefont {Jungmann}}, \bibinfo
  {author} {\bibfnamefont {D.}~\bibnamefont {Wolf}}, \bibinfo {author}
  {\bibfnamefont {L.}~\bibnamefont {Brendel}}, \bibinfo {author} {\bibfnamefont
  {J.}~\bibnamefont {Teiser}}, \bibinfo {author} {\bibfnamefont
  {T.}~\bibnamefont {Shinbrot}},\ and\ \bibinfo {author} {\bibfnamefont
  {G.}~\bibnamefont {Wurm}},\ }\href
  {https://doi.org/10.1038/s41567-019-0728-9} {\bibfield  {journal} {\bibinfo
  {journal} {Nat. Phys.}\ }\textbf {\bibinfo {volume} {16}},\ \bibinfo {pages}
  {225} (\bibinfo {year} {2020})}\BibitemShut {NoStop}%
\bibitem [{\citenamefont {Singh}\ and\ \citenamefont
  {Mazza}(2018)}]{Singh.2018}%
  \BibitemOpen
  \bibfield  {author} {\bibinfo {author} {\bibfnamefont {C.}~\bibnamefont
  {Singh}}\ and\ \bibinfo {author} {\bibfnamefont {M.~G.}\ \bibnamefont
  {Mazza}},\ }\href {https://doi.org/10.1103/PhysRevE.97.022904} {\bibfield
  {journal} {\bibinfo  {journal} {Phys. Rev. E}\ }\textbf {\bibinfo {volume}
  {97}},\ \bibinfo {pages} {022904} (\bibinfo {year} {2018})}\BibitemShut
  {NoStop}%
\bibitem [{\citenamefont {Lee}\ \emph {et~al.}(2015)\citenamefont {Lee},
  \citenamefont {Waitukaitis}, \citenamefont {Miskin},\ and\ \citenamefont
  {Jaeger}}]{Lee.2015}%
  \BibitemOpen
  \bibfield  {author} {\bibinfo {author} {\bibfnamefont {V.}~\bibnamefont
  {Lee}}, \bibinfo {author} {\bibfnamefont {S.~R.}\ \bibnamefont
  {Waitukaitis}}, \bibinfo {author} {\bibfnamefont {M.~Z.}\ \bibnamefont
  {Miskin}},\ and\ \bibinfo {author} {\bibfnamefont {H.~M.}\ \bibnamefont
  {Jaeger}},\ }\href {https://doi.org/10.1038/nphys3396} {\bibfield  {journal}
  {\bibinfo  {journal} {Nat. Phys.}\ }\textbf {\bibinfo {volume} {11}},\
  \bibinfo {pages} {733 } (\bibinfo {year} {2015})}\BibitemShut {NoStop}%
\bibitem [{\citenamefont {Matsusyama}\ and\ \citenamefont
  {Yamamoto}(2006)}]{Matsuyama.2006}%
  \BibitemOpen
  \bibfield  {author} {\bibinfo {author} {\bibfnamefont {T.}~\bibnamefont
  {Matsusyama}}\ and\ \bibinfo {author} {\bibfnamefont {H.}~\bibnamefont
  {Yamamoto}},\ }\href {https://doi.org/10.1016/j.ces.2005.05.003} {\bibfield
  {journal} {\bibinfo  {journal} {Chem. Eng. Sci.}\ }\textbf {\bibinfo {volume}
  {61}},\ \bibinfo {pages} {2230} (\bibinfo {year} {2006})}\BibitemShut
  {NoStop}%
\bibitem [{\citenamefont {Ireland}(2010)}]{Ireland.2010}%
  \BibitemOpen
  \bibfield  {author} {\bibinfo {author} {\bibfnamefont {P.~M.}\ \bibnamefont
  {Ireland}},\ }\href {https://doi.org/10.1016/j.powtec.2009.11.017} {\bibfield
   {journal} {\bibinfo  {journal} {Powder Technol.}\ }\textbf {\bibinfo
  {volume} {198}},\ \bibinfo {pages} {189} (\bibinfo {year}
  {2010})}\BibitemShut {NoStop}%
\bibitem [{\citenamefont {Grosshans}\ and\ \citenamefont
  {Papalexandris}(2017)}]{Grosshans.2017}%
  \BibitemOpen
  \bibfield  {author} {\bibinfo {author} {\bibfnamefont {H.}~\bibnamefont
  {Grosshans}}\ and\ \bibinfo {author} {\bibfnamefont {M.~V.}\ \bibnamefont
  {Papalexandris}},\ }\href {https://doi.org/10.1016/j.powtec.2016.10.024}
  {\bibfield  {journal} {\bibinfo  {journal} {Powder Technol.}\ }\textbf
  {\bibinfo {volume} {305}},\ \bibinfo {pages} {518} (\bibinfo {year}
  {2017})}\BibitemShut {NoStop}%
\bibitem [{\citenamefont {Lowell}\ and\ \citenamefont
  {Truscott}(1986)}]{Lowell.1986b}%
  \BibitemOpen
  \bibfield  {author} {\bibinfo {author} {\bibfnamefont {J.}~\bibnamefont
  {Lowell}}\ and\ \bibinfo {author} {\bibfnamefont {W.}~\bibnamefont
  {Truscott}},\ }\href@noop {} {\bibfield  {journal} {\bibinfo  {journal} {J.
  Phys. D}\ }\textbf {\bibinfo {volume} {19}},\ \bibinfo {pages} {1281}
  (\bibinfo {year} {1986})}\BibitemShut {NoStop}%
\bibitem [{\citenamefont {Lacks}\ \emph {et~al.}(2008)\citenamefont {Lacks},
  \citenamefont {Duff},\ and\ \citenamefont {Kumar}}]{Lacks.2008}%
  \BibitemOpen
  \bibfield  {author} {\bibinfo {author} {\bibfnamefont {D.~J.}\ \bibnamefont
  {Lacks}}, \bibinfo {author} {\bibfnamefont {N.}~\bibnamefont {Duff}},\ and\
  \bibinfo {author} {\bibfnamefont {S.~K.}\ \bibnamefont {Kumar}},\ }\href
  {https://doi.org/10.1103/physrevlett.100.188305} {\bibfield  {journal}
  {\bibinfo  {journal} {Phys. Rev. Lett.}\ }\textbf {\bibinfo {volume} {100}},\
  \bibinfo {pages} {188305} (\bibinfo {year} {2008})}\BibitemShut {NoStop}%
\bibitem [{\citenamefont {Kok}\ and\ \citenamefont {Lacks}(2009)}]{Kok.2009}%
  \BibitemOpen
  \bibfield  {author} {\bibinfo {author} {\bibfnamefont {J.~F.}\ \bibnamefont
  {Kok}}\ and\ \bibinfo {author} {\bibfnamefont {D.~J.}\ \bibnamefont
  {Lacks}},\ }\href {https://doi.org/10.1103/physreve.79.051304} {\bibfield
  {journal} {\bibinfo  {journal} {Phys. Rev. E}\ }\textbf {\bibinfo {volume}
  {79}},\ \bibinfo {pages} {051304} (\bibinfo {year} {2009})}\BibitemShut
  {NoStop}%
\bibitem [{\citenamefont {Forward}\ \emph {et~al.}(2009)\citenamefont
  {Forward}, \citenamefont {Lacks},\ and\ \citenamefont
  {Sankaran}}]{Forward.2009bs}%
  \BibitemOpen
  \bibfield  {author} {\bibinfo {author} {\bibfnamefont {K.~M.}\ \bibnamefont
  {Forward}}, \bibinfo {author} {\bibfnamefont {D.~J.}\ \bibnamefont {Lacks}},\
  and\ \bibinfo {author} {\bibfnamefont {R.~M.}\ \bibnamefont {Sankaran}},\
  }\href {https://doi.org/10.1103/physrevlett.102.028001} {\bibfield  {journal}
  {\bibinfo  {journal} {Phys. Rev. Lett.}\ }\textbf {\bibinfo {volume} {102}},\
  \bibinfo {pages} {028001} (\bibinfo {year} {2009})}\BibitemShut {NoStop}%
\bibitem [{\citenamefont {Apodaca}\ \emph {et~al.}(2010)\citenamefont
  {Apodaca}, \citenamefont {Wesson}, \citenamefont {Bishop}, \citenamefont
  {Ratner},\ and\ \citenamefont {Grzybowski}}]{apodaca2010}%
  \BibitemOpen
  \bibfield  {author} {\bibinfo {author} {\bibfnamefont {M.~M.}\ \bibnamefont
  {Apodaca}}, \bibinfo {author} {\bibfnamefont {P.~J.}\ \bibnamefont {Wesson}},
  \bibinfo {author} {\bibfnamefont {K.~J.~M.}\ \bibnamefont {Bishop}}, \bibinfo
  {author} {\bibfnamefont {M.~A.}\ \bibnamefont {Ratner}},\ and\ \bibinfo
  {author} {\bibfnamefont {B.~A.}\ \bibnamefont {Grzybowski}},\ }\href
  {https://doi.org/10.1002/anie.200905281} {\bibfield  {journal} {\bibinfo
  {journal} {Angew. Chem., Int. Ed. Engl.}\ }\textbf {\bibinfo {volume} {49}},\
  \bibinfo {pages} {946} (\bibinfo {year} {2010})}\BibitemShut {NoStop}%
\bibitem [{\citenamefont {Yu}\ \emph {et~al.}(2017)\citenamefont {Yu},
  \citenamefont {Mu},\ and\ \citenamefont {Xie}}]{Yu.2017}%
  \BibitemOpen
  \bibfield  {author} {\bibinfo {author} {\bibfnamefont {H.}~\bibnamefont
  {Yu}}, \bibinfo {author} {\bibfnamefont {L.}~\bibnamefont {Mu}},\ and\
  \bibinfo {author} {\bibfnamefont {L.}~\bibnamefont {Xie}},\ }\href
  {https://doi.org/10.1016/j.elstat.2017.10.001} {\bibfield  {journal}
  {\bibinfo  {journal} {J. Electrostat.}\ }\textbf {\bibinfo {volume} {90}},\
  \bibinfo {pages} {113 } (\bibinfo {year} {2017})}\BibitemShut {NoStop}%
\bibitem [{\citenamefont {Xie}\ \emph {et~al.}(2016)\citenamefont {Xie},
  \citenamefont {Bao}, \citenamefont {Jiang},\ and\ \citenamefont
  {Zhou}}]{xie2016}%
  \BibitemOpen
  \bibfield  {author} {\bibinfo {author} {\bibfnamefont {L.}~\bibnamefont
  {Xie}}, \bibinfo {author} {\bibfnamefont {N.}~\bibnamefont {Bao}}, \bibinfo
  {author} {\bibfnamefont {Y.}~\bibnamefont {Jiang}},\ and\ \bibinfo {author}
  {\bibfnamefont {J.}~\bibnamefont {Zhou}},\ }\href
  {https://doi.org/10.1063/1.4944831} {\bibfield  {journal} {\bibinfo
  {journal} {AIP Adv.}\ }\textbf {\bibinfo {volume} {6}},\ \bibinfo {pages}
  {035117} (\bibinfo {year} {2016})}\BibitemShut {NoStop}%
\bibitem [{\citenamefont {Grosjean}\ \emph {et~al.}(2020)\citenamefont
  {Grosjean}, \citenamefont {Wald}, \citenamefont {Sobarzo},\ and\
  \citenamefont {Waitukaitis}}]{grosjean2020}%
  \BibitemOpen
  \bibfield  {author} {\bibinfo {author} {\bibfnamefont {G.}~\bibnamefont
  {Grosjean}}, \bibinfo {author} {\bibfnamefont {S.}~\bibnamefont {Wald}},
  \bibinfo {author} {\bibfnamefont {J.~C.}\ \bibnamefont {Sobarzo}},\ and\
  \bibinfo {author} {\bibfnamefont {S.}~\bibnamefont {Waitukaitis}},\ }\href
  {https://doi.org/10.1103/PhysRevMaterials.4.082602} {\bibfield  {journal}
  {\bibinfo  {journal} {Phys. Rev. Mater.}\ }\textbf {\bibinfo {volume} {4}},\
  \bibinfo {pages} {082602(R)} (\bibinfo {year} {2020})}\BibitemShut {NoStop}%
\bibitem [{\citenamefont {Zheng}\ \emph {et~al.}(2014)\citenamefont {Zheng},
  \citenamefont {Zhang},\ and\ \citenamefont {Huang}}]{Zheng.2014}%
  \BibitemOpen
  \bibfield  {author} {\bibinfo {author} {\bibfnamefont {X.}~\bibnamefont
  {Zheng}}, \bibinfo {author} {\bibfnamefont {R.}~\bibnamefont {Zhang}},\ and\
  \bibinfo {author} {\bibfnamefont {H.}~\bibnamefont {Huang}},\ }\href
  {https://doi.org/10.1038/srep04399} {\bibfield  {journal} {\bibinfo
  {journal} {Sci. Rep.}\ }\textbf {\bibinfo {volume} {4}},\ \bibinfo {pages}
  {4399} (\bibinfo {year} {2014})}\BibitemShut {NoStop}%
\bibitem [{\citenamefont {Zhang}\ \emph {et~al.}(2015)\citenamefont {Zhang},
  \citenamefont {P\"ahtz}, \citenamefont {Liu}, \citenamefont {Wang},
  \citenamefont {Zhang}, \citenamefont {Shen}, \citenamefont {Ji},\ and\
  \citenamefont {Cai}}]{Zhang.2015}%
  \BibitemOpen
  \bibfield  {author} {\bibinfo {author} {\bibfnamefont {Y.}~\bibnamefont
  {Zhang}}, \bibinfo {author} {\bibfnamefont {T.}~\bibnamefont {P\"ahtz}},
  \bibinfo {author} {\bibfnamefont {Y.}~\bibnamefont {Liu}}, \bibinfo {author}
  {\bibfnamefont {X.}~\bibnamefont {Wang}}, \bibinfo {author} {\bibfnamefont
  {R.}~\bibnamefont {Zhang}}, \bibinfo {author} {\bibfnamefont
  {Y.}~\bibnamefont {Shen}}, \bibinfo {author} {\bibfnamefont {R.}~\bibnamefont
  {Ji}},\ and\ \bibinfo {author} {\bibfnamefont {B.}~\bibnamefont {Cai}},\
  }\href {https://doi.org/10.1103/physrevx.5.011002} {\bibfield  {journal}
  {\bibinfo  {journal} {Phys. Rev. X}\ }\textbf {\bibinfo {volume} {5}},\
  \bibinfo {pages} {011002} (\bibinfo {year} {2015})}\BibitemShut {NoStop}%
\bibitem [{\citenamefont {Lee}\ \emph {et~al.}(2018)\citenamefont {Lee},
  \citenamefont {James}, \citenamefont {Waitukaitis},\ and\ \citenamefont
  {Jaeger}}]{lee2018}%
  \BibitemOpen
  \bibfield  {author} {\bibinfo {author} {\bibfnamefont {V.}~\bibnamefont
  {Lee}}, \bibinfo {author} {\bibfnamefont {N.~M.}\ \bibnamefont {James}},
  \bibinfo {author} {\bibfnamefont {S.~R.}\ \bibnamefont {Waitukaitis}},\ and\
  \bibinfo {author} {\bibfnamefont {H.~M.}\ \bibnamefont {Jaeger}},\ }\href
  {https://doi.org/10.1103/PhysRevMaterials.2.035602} {\bibfield  {journal}
  {\bibinfo  {journal} {Phys. Rev. Mater.}\ }\textbf {\bibinfo {volume} {2}},\
  \bibinfo {pages} {035602} (\bibinfo {year} {2018})}\BibitemShut {NoStop}%
\bibitem [{\citenamefont {LaMarche}\ \emph {et~al.}(2009)\citenamefont
  {LaMarche}, \citenamefont {Liu}, \citenamefont {Shah}, \citenamefont
  {Shinbrot},\ and\ \citenamefont {Glasser}}]{LaMarche.2009}%
  \BibitemOpen
  \bibfield  {author} {\bibinfo {author} {\bibfnamefont {K.~R.}\ \bibnamefont
  {LaMarche}}, \bibinfo {author} {\bibfnamefont {X.}~\bibnamefont {Liu}},
  \bibinfo {author} {\bibfnamefont {S.~K.}\ \bibnamefont {Shah}}, \bibinfo
  {author} {\bibfnamefont {T.}~\bibnamefont {Shinbrot}},\ and\ \bibinfo
  {author} {\bibfnamefont {B.~J.}\ \bibnamefont {Glasser}},\ }\href
  {https://doi.org/10.1016/j.powtec.2009.05.026} {\bibfield  {journal}
  {\bibinfo  {journal} {Powder Technol.}\ }\textbf {\bibinfo {volume} {195}},\
  \bibinfo {pages} {158 } (\bibinfo {year} {2009})}\BibitemShut {NoStop}%
\bibitem [{\citenamefont {Waitukaitis}\ and\ \citenamefont
  {Jaeger}(2013)}]{Waitukaitis.2013un5}%
  \BibitemOpen
  \bibfield  {author} {\bibinfo {author} {\bibfnamefont {S.~R.}\ \bibnamefont
  {Waitukaitis}}\ and\ \bibinfo {author} {\bibfnamefont {H.~M.}\ \bibnamefont
  {Jaeger}},\ }\href {https://doi.org/10.1063/1.4789496} {\bibfield  {journal}
  {\bibinfo  {journal} {Rev. Sci. Instrum.}\ }\textbf {\bibinfo {volume}
  {84}},\ \bibinfo {pages} {025104 } (\bibinfo {year} {2013})}\BibitemShut
  {NoStop}%
\bibitem [{\citenamefont {Waitukaitis}\ \emph {et~al.}(2014)\citenamefont
  {Waitukaitis}, \citenamefont {Lee}, \citenamefont {Pierson}, \citenamefont
  {Forman},\ and\ \citenamefont {Jaeger}}]{Waitukaitis.2014}%
  \BibitemOpen
  \bibfield  {author} {\bibinfo {author} {\bibfnamefont {S.~R.}\ \bibnamefont
  {Waitukaitis}}, \bibinfo {author} {\bibfnamefont {V.}~\bibnamefont {Lee}},
  \bibinfo {author} {\bibfnamefont {J.~M.}\ \bibnamefont {Pierson}}, \bibinfo
  {author} {\bibfnamefont {S.~L.}\ \bibnamefont {Forman}},\ and\ \bibinfo
  {author} {\bibfnamefont {H.~M.}\ \bibnamefont {Jaeger}},\ }\href
  {https://doi.org/10.1103/physrevlett.112.218001} {\bibfield  {journal}
  {\bibinfo  {journal} {Phys. Rev. Lett.}\ }\textbf {\bibinfo {volume} {112}},\
  \bibinfo {pages} {218001 } (\bibinfo {year} {2014})}\BibitemShut {NoStop}%
\bibitem [{\citenamefont {Carter}\ and\ \citenamefont
  {Hartzell}(2020)}]{Carter.2020}%
  \BibitemOpen
  \bibfield  {author} {\bibinfo {author} {\bibfnamefont {D.}~\bibnamefont
  {Carter}}\ and\ \bibinfo {author} {\bibfnamefont {C.}~\bibnamefont
  {Hartzell}},\ }\href {https://doi.org/10.1016/j.elstat.2020.103475}
  {\bibfield  {journal} {\bibinfo  {journal} {J. Electrostat.}\ }\textbf
  {\bibinfo {volume} {107}},\ \bibinfo {pages} {103475} (\bibinfo {year}
  {2020})}\BibitemShut {NoStop}%
\bibitem [{\citenamefont {Matsuyama}\ and\ \citenamefont
  {Yamamoto}(1995)}]{Matsuyama.1995}%
  \BibitemOpen
  \bibfield  {author} {\bibinfo {author} {\bibfnamefont {T.}~\bibnamefont
  {Matsuyama}}\ and\ \bibinfo {author} {\bibfnamefont {H.}~\bibnamefont
  {Yamamoto}},\ }\href {https://doi.org/10.1088/0022-3727/28/12/005} {\bibfield
   {journal} {\bibinfo  {journal} {J. Phys. D}\ }\textbf {\bibinfo {volume}
  {28}},\ \bibinfo {pages} {2418} (\bibinfo {year} {1995})}\BibitemShut
  {NoStop}%
\bibitem [{\citenamefont {Matsuyama}\ \emph {et~al.}(2003)\citenamefont
  {Matsuyama}, \citenamefont {Ogu}, \citenamefont {Yamamoto}, \citenamefont
  {Marijnissen},\ and\ \citenamefont {Scarlett}}]{Matsuyama.2003}%
  \BibitemOpen
  \bibfield  {author} {\bibinfo {author} {\bibfnamefont {T.}~\bibnamefont
  {Matsuyama}}, \bibinfo {author} {\bibfnamefont {M.}~\bibnamefont {Ogu}},
  \bibinfo {author} {\bibfnamefont {H.}~\bibnamefont {Yamamoto}}, \bibinfo
  {author} {\bibfnamefont {J.~C.}\ \bibnamefont {Marijnissen}},\ and\ \bibinfo
  {author} {\bibfnamefont {B.}~\bibnamefont {Scarlett}},\ }\href
  {https://doi.org/10.1016/s0032-5910(03)00154-2} {\bibfield  {journal}
  {\bibinfo  {journal} {Powder Technol.}\ }\textbf {\bibinfo {volume} {135}},\
  \bibinfo {pages} {14} (\bibinfo {year} {2003})}\BibitemShut {NoStop}%
\bibitem [{\citenamefont {Watanabe}\ \emph {et~al.}(2006)\citenamefont
  {Watanabe}, \citenamefont {Samimi}, \citenamefont {Ding}, \citenamefont
  {Ghadiri}, \citenamefont {Matsuyama},\ and\ \citenamefont
  {Pitt}}]{Watanabe.2006}%
  \BibitemOpen
  \bibfield  {author} {\bibinfo {author} {\bibfnamefont {H.}~\bibnamefont
  {Watanabe}}, \bibinfo {author} {\bibfnamefont {A.}~\bibnamefont {Samimi}},
  \bibinfo {author} {\bibfnamefont {Y.~L.}\ \bibnamefont {Ding}}, \bibinfo
  {author} {\bibfnamefont {M.}~\bibnamefont {Ghadiri}}, \bibinfo {author}
  {\bibfnamefont {T.}~\bibnamefont {Matsuyama}},\ and\ \bibinfo {author}
  {\bibfnamefont {K.~G.}\ \bibnamefont {Pitt}},\ }\href
  {https://doi.org/10.1002/ppsc.200601021} {\bibfield  {journal} {\bibinfo
  {journal} {Part. Part. Syst. Charact.}\ }\textbf {\bibinfo {volume} {23}},\
  \bibinfo {pages} {133} (\bibinfo {year} {2006})}\BibitemShut {NoStop}%
\bibitem [{\citenamefont {Haeberle}\ \emph {et~al.}(2018)\citenamefont
  {Haeberle}, \citenamefont {Schella}, \citenamefont {Sperl}, \citenamefont
  {Schröter},\ and\ \citenamefont {Born}}]{Haeberle.2018}%
  \BibitemOpen
  \bibfield  {author} {\bibinfo {author} {\bibfnamefont {J.}~\bibnamefont
  {Haeberle}}, \bibinfo {author} {\bibfnamefont {A.}~\bibnamefont {Schella}},
  \bibinfo {author} {\bibfnamefont {M.}~\bibnamefont {Sperl}}, \bibinfo
  {author} {\bibfnamefont {M.}~\bibnamefont {Schröter}},\ and\ \bibinfo
  {author} {\bibfnamefont {P.}~\bibnamefont {Born}},\ }\href
  {https://doi.org/10.1039/c8sm00603b} {\bibfield  {journal} {\bibinfo
  {journal} {Soft Matter}\ }\textbf {\bibinfo {volume} {14}},\ \bibinfo {pages}
  {4987 } (\bibinfo {year} {2018})}\BibitemShut {NoStop}%
\bibitem [{\citenamefont {Xie}\ \emph {et~al.}(2013)\citenamefont {Xie},
  \citenamefont {Li}, \citenamefont {Bao},\ and\ \citenamefont
  {Zhou}}]{Xie.2013}%
  \BibitemOpen
  \bibfield  {author} {\bibinfo {author} {\bibfnamefont {L.}~\bibnamefont
  {Xie}}, \bibinfo {author} {\bibfnamefont {G.}~\bibnamefont {Li}}, \bibinfo
  {author} {\bibfnamefont {N.}~\bibnamefont {Bao}},\ and\ \bibinfo {author}
  {\bibfnamefont {J.}~\bibnamefont {Zhou}},\ }\href
  {https://doi.org/10.1063/1.4804331} {\bibfield  {journal} {\bibinfo
  {journal} {J. Appl. Phys.}\ }\textbf {\bibinfo {volume} {113}},\ \bibinfo
  {pages} {184908 } (\bibinfo {year} {2013})}\BibitemShut {NoStop}%
\bibitem [{\citenamefont {Hu}\ \emph {et~al.}(2012)\citenamefont {Hu},
  \citenamefont {Xie},\ and\ \citenamefont {Zheng}}]{Hu.2012}%
  \BibitemOpen
  \bibfield  {author} {\bibinfo {author} {\bibfnamefont {W.}~\bibnamefont
  {Hu}}, \bibinfo {author} {\bibfnamefont {L.}~\bibnamefont {Xie}},\ and\
  \bibinfo {author} {\bibfnamefont {X.}~\bibnamefont {Zheng}},\ }\href
  {https://doi.org/10.1063/1.4752458} {\bibfield  {journal} {\bibinfo
  {journal} {Appl. Phys. Lett.}\ }\textbf {\bibinfo {volume} {101}},\ \bibinfo
  {pages} {114107 } (\bibinfo {year} {2012})}\BibitemShut {NoStop}%
\bibitem [{\citenamefont {Kline}\ \emph {et~al.}(2020)\citenamefont {Kline},
  \citenamefont {Lim},\ and\ \citenamefont {Jaeger}}]{kline2020}%
  \BibitemOpen
  \bibfield  {author} {\bibinfo {author} {\bibfnamefont {A.~G.}\ \bibnamefont
  {Kline}}, \bibinfo {author} {\bibfnamefont {M.~X.}\ \bibnamefont {Lim}},\
  and\ \bibinfo {author} {\bibfnamefont {H.~M.}\ \bibnamefont {Jaeger}},\
  }\href {https://doi.org/10.1063/1.5133049} {\bibfield  {journal} {\bibinfo
  {journal} {Rev. Sci. Instrum.}\ }\textbf {\bibinfo {volume} {91}},\ \bibinfo
  {pages} {023908} (\bibinfo {year} {2020})}\BibitemShut {NoStop}%
\bibitem [{Sup()}]{SupplMat}%
  \BibitemOpen
  \href@noop {} {\bibinfo  {journal} {See Supplemental Material at
  \url{https://arxiv.org/src/2211.02488v4/anc} for the supplemental videos and
  further experimental details.}\ }\BibitemShut {NoStop}%
\bibitem [{\citenamefont {Andrade}\ \emph {et~al.}(2018)\citenamefont
  {Andrade}, \citenamefont {P{\'e}rez},\ and\ \citenamefont
  {Adamowski}}]{andrade2018}%
  \BibitemOpen
\bibfield  {journal} {  }\bibfield  {author} {\bibinfo {author} {\bibfnamefont
  {M.~A.~B.}\ \bibnamefont {Andrade}}, \bibinfo {author} {\bibfnamefont
  {N.}~\bibnamefont {P{\'e}rez}},\ and\ \bibinfo {author} {\bibfnamefont
  {J.~C.}\ \bibnamefont {Adamowski}},\ }\href
  {https://doi.org/10.1007/s13538-017-0552-6} {\bibfield  {journal} {\bibinfo
  {journal} {Braz. J. Phys.}\ }\textbf {\bibinfo {volume} {48}},\ \bibinfo
  {pages} {190} (\bibinfo {year} {2018})}\BibitemShut {NoStop}%
\bibitem [{\citenamefont {Harper}\ \emph {et~al.}(2022)\citenamefont {Harper},
  \citenamefont {Harvey}, \citenamefont {Huang}, \citenamefont {McGrath},
  \citenamefont {Meer},\ and\ \citenamefont {Burton}}]{Harvey.2022}%
  \BibitemOpen
  \bibfield  {author} {\bibinfo {author} {\bibfnamefont {J.~M.}\ \bibnamefont
  {Harper}}, \bibinfo {author} {\bibfnamefont {D.}~\bibnamefont {Harvey}},
  \bibinfo {author} {\bibfnamefont {T.}~\bibnamefont {Huang}}, \bibinfo
  {author} {\bibfnamefont {J.}~\bibnamefont {McGrath}}, \bibinfo {author}
  {\bibfnamefont {D.}~\bibnamefont {Meer}},\ and\ \bibinfo {author}
  {\bibfnamefont {J.~C.}\ \bibnamefont {Burton}},\ }\href
  {https://doi.org/10.1093/pnasnexus/pgac220} {\bibfield  {journal} {\bibinfo
  {journal} {{PNAS} Nexus}\ }\textbf {\bibinfo {volume} {1}},\ \bibinfo {pages}
  {pgac220} (\bibinfo {year} {2022})}\BibitemShut {NoStop}%
\bibitem [{\citenamefont {Yoshimatsu}\ \emph {et~al.}(2017)\citenamefont
  {Yoshimatsu}, \citenamefont {Araújo}, \citenamefont {Wurm}, \citenamefont
  {Herrmann},\ and\ \citenamefont {Shinbrot}}]{yoshimatsu2017}%
  \BibitemOpen
  \bibfield  {author} {\bibinfo {author} {\bibfnamefont {R.}~\bibnamefont
  {Yoshimatsu}}, \bibinfo {author} {\bibfnamefont {N.~a.~M.}\ \bibnamefont
  {Araújo}}, \bibinfo {author} {\bibfnamefont {G.}~\bibnamefont {Wurm}},
  \bibinfo {author} {\bibfnamefont {H.~J.}\ \bibnamefont {Herrmann}},\ and\
  \bibinfo {author} {\bibfnamefont {T.}~\bibnamefont {Shinbrot}},\ }\href
  {https://doi.org/10.1038/srep39996} {\bibfield  {journal} {\bibinfo
  {journal} {Sci. Rep.}\ }\textbf {\bibinfo {volume} {7}},\ \bibinfo {pages}
  {39996} (\bibinfo {year} {2017})}\BibitemShut {NoStop}%
\bibitem [{\citenamefont {Zhuravlev}(2000)}]{Zhuravlev.2000}%
  \BibitemOpen
  \bibfield  {author} {\bibinfo {author} {\bibfnamefont {L.}~\bibnamefont
  {Zhuravlev}},\ }\href {https://doi.org/10.1016/s0927-7757(00)00556-2}
  {\bibfield  {journal} {\bibinfo  {journal} {Colloids Surf. A}\ }\textbf
  {\bibinfo {volume} {173}},\ \bibinfo {pages} {1} (\bibinfo {year}
  {2000})}\BibitemShut {NoStop}%
\bibitem [{\citenamefont {Burgo}\ \emph {et~al.}(2016)\citenamefont {Burgo},
  \citenamefont {Galembeck},\ and\ \citenamefont {Pollack}}]{Burgo.2016}%
  \BibitemOpen
  \bibfield  {author} {\bibinfo {author} {\bibfnamefont {T.~A.}\ \bibnamefont
  {Burgo}}, \bibinfo {author} {\bibfnamefont {F.}~\bibnamefont {Galembeck}},\
  and\ \bibinfo {author} {\bibfnamefont {G.~H.}\ \bibnamefont {Pollack}},\
  }\href {https://doi.org/10.1016/j.elstat.2016.01.002} {\bibfield  {journal}
  {\bibinfo  {journal} {J. Electrostat.}\ }\textbf {\bibinfo {volume} {80}},\
  \bibinfo {pages} {30} (\bibinfo {year} {2016})}\BibitemShut {NoStop}%
\bibitem [{\citenamefont {Gil}\ and\ \citenamefont {Lacks}(2019)}]{Gil.2019}%
  \BibitemOpen
  \bibfield  {author} {\bibinfo {author} {\bibfnamefont {P.~S.}\ \bibnamefont
  {Gil}}\ and\ \bibinfo {author} {\bibfnamefont {D.~J.}\ \bibnamefont
  {Lacks}},\ }\href {https://doi.org/10.1039/c9cp02398d} {\bibfield  {journal}
  {\bibinfo  {journal} {Phys. Chem. Chem. Phys.}\ }\textbf {\bibinfo {volume}
  {21}},\ \bibinfo {pages} {13821 } (\bibinfo {year} {2019})}\BibitemShut
  {NoStop}%
\bibitem [{\citenamefont {Jungmann}\ \emph {et~al.}(2022)\citenamefont
  {Jungmann}, \citenamefont {Onyeagusi}, \citenamefont {Teiser},\ and\
  \citenamefont {Wurm}}]{Jungmann.2022x}%
  \BibitemOpen
  \bibfield  {author} {\bibinfo {author} {\bibfnamefont {F.}~\bibnamefont
  {Jungmann}}, \bibinfo {author} {\bibfnamefont {F.~C.}\ \bibnamefont
  {Onyeagusi}}, \bibinfo {author} {\bibfnamefont {J.}~\bibnamefont {Teiser}},\
  and\ \bibinfo {author} {\bibfnamefont {G.}~\bibnamefont {Wurm}},\ }\href
  {https://doi.org/10.1016/j.elstat.2022.103705} {\bibfield  {journal}
  {\bibinfo  {journal} {J. Electrostat.}\ }\textbf {\bibinfo {volume} {117}},\
  \bibinfo {pages} {103705} (\bibinfo {year} {2022})}\BibitemShut {NoStop}%
\bibitem [{\citenamefont {Schella}\ \emph {et~al.}(2017)\citenamefont
  {Schella}, \citenamefont {Herminghaus},\ and\ \citenamefont
  {Schröter}}]{schella2017}%
  \BibitemOpen
  \bibfield  {author} {\bibinfo {author} {\bibfnamefont {A.}~\bibnamefont
  {Schella}}, \bibinfo {author} {\bibfnamefont {S.}~\bibnamefont
  {Herminghaus}},\ and\ \bibinfo {author} {\bibfnamefont {M.}~\bibnamefont
  {Schröter}},\ }\href {https://doi.org/10.1039/C6SM02041K} {\bibfield
  {journal} {\bibinfo  {journal} {Soft Matter}\ }\textbf {\bibinfo {volume}
  {13}},\ \bibinfo {pages} {394} (\bibinfo {year} {2017})}\BibitemShut
  {NoStop}%
\bibitem [{\citenamefont {Sullivan}\ \emph {et~al.}(2007)\citenamefont
  {Sullivan}, \citenamefont {Stone}, \citenamefont {Hashisho},\ and\
  \citenamefont {Rood}}]{Sullivan.2007cat}%
  \BibitemOpen
  \bibfield  {author} {\bibinfo {author} {\bibfnamefont {P.~D.}\ \bibnamefont
  {Sullivan}}, \bibinfo {author} {\bibfnamefont {B.~R.}\ \bibnamefont {Stone}},
  \bibinfo {author} {\bibfnamefont {Z.}~\bibnamefont {Hashisho}},\ and\
  \bibinfo {author} {\bibfnamefont {M.~J.}\ \bibnamefont {Rood}},\ }\href
  {https://doi.org/10.1007/s10450-007-9033-5} {\bibfield  {journal} {\bibinfo
  {journal} {Adsorption}\ }\textbf {\bibinfo {volume} {13}},\ \bibinfo {pages}
  {173} (\bibinfo {year} {2007})}\BibitemShut {NoStop}%
\bibitem [{\citenamefont {Taqvi}\ \emph {et~al.}(1999)\citenamefont {Taqvi},
  \citenamefont {Appel},\ and\ \citenamefont {LeVan}}]{Taqvi.1999}%
  \BibitemOpen
  \bibfield  {author} {\bibinfo {author} {\bibfnamefont {S.~M.}\ \bibnamefont
  {Taqvi}}, \bibinfo {author} {\bibfnamefont {W.~S.}\ \bibnamefont {Appel}},\
  and\ \bibinfo {author} {\bibfnamefont {M.~D.}\ \bibnamefont {LeVan}},\ }\href
  {https://doi.org/10.1021/ie980324k} {\bibfield  {journal} {\bibinfo
  {journal} {Ind. Eng. Chem. Res.}\ }\textbf {\bibinfo {volume} {38}},\
  \bibinfo {pages} {240} (\bibinfo {year} {1999})}\BibitemShut {NoStop}%
\bibitem [{\citenamefont {Rudisill}\ \emph {et~al.}(1992)\citenamefont
  {Rudisill}, \citenamefont {Hacskaylo},\ and\ \citenamefont
  {LeVan}}]{Rudisill.1992}%
  \BibitemOpen
  \bibfield  {author} {\bibinfo {author} {\bibfnamefont {E.~N.}\ \bibnamefont
  {Rudisill}}, \bibinfo {author} {\bibfnamefont {J.~J.}\ \bibnamefont
  {Hacskaylo}},\ and\ \bibinfo {author} {\bibfnamefont {M.~D.}\ \bibnamefont
  {LeVan}},\ }\href {https://doi.org/https://doi.org/10.1021/ie00004a022}
  {\bibfield  {journal} {\bibinfo  {journal} {Ind. Eng. Chem. Res.}\ }\textbf
  {\bibinfo {volume} {31}},\ \bibinfo {pages} {1122} (\bibinfo {year}
  {1992})}\BibitemShut {NoStop}%
\bibitem [{\citenamefont {Barnette}\ and\ \citenamefont
  {Kim}(2012)}]{Barnette.2012}%
  \BibitemOpen
  \bibfield  {author} {\bibinfo {author} {\bibfnamefont {A.~L.}\ \bibnamefont
  {Barnette}}\ and\ \bibinfo {author} {\bibfnamefont {S.~H.}\ \bibnamefont
  {Kim}},\ }\href {https://doi.org/10.1021/la302849t} {\bibfield  {journal}
  {\bibinfo  {journal} {Langmuir}\ }\textbf {\bibinfo {volume} {28}},\ \bibinfo
  {pages} {15529} (\bibinfo {year} {2012})}\BibitemShut {NoStop}%
\bibitem [{\citenamefont {Qi}\ and\ \citenamefont {LeVan}(2005)}]{Qi.2005abc}%
  \BibitemOpen
  \bibfield  {author} {\bibinfo {author} {\bibfnamefont {N.}~\bibnamefont
  {Qi}}\ and\ \bibinfo {author} {\bibfnamefont {M.~D.}\ \bibnamefont {LeVan}},\
  }\href {https://doi.org/10.1021/ie049109w} {\bibfield  {journal} {\bibinfo
  {journal} {Ind. Eng. Chem. Res.}\ }\textbf {\bibinfo {volume} {44}},\
  \bibinfo {pages} {3733} (\bibinfo {year} {2005})}\BibitemShut {NoStop}%
\bibitem [{\citenamefont {Kierlik}\ \emph {et~al.}(2001)\citenamefont
  {Kierlik}, \citenamefont {Monson}, \citenamefont {Rosinberg}, \citenamefont
  {Sarkisov},\ and\ \citenamefont {Tarjus}}]{kierlik2001}%
  \BibitemOpen
  \bibfield  {author} {\bibinfo {author} {\bibfnamefont {E.}~\bibnamefont
  {Kierlik}}, \bibinfo {author} {\bibfnamefont {P.~A.}\ \bibnamefont {Monson}},
  \bibinfo {author} {\bibfnamefont {M.~L.}\ \bibnamefont {Rosinberg}}, \bibinfo
  {author} {\bibfnamefont {L.}~\bibnamefont {Sarkisov}},\ and\ \bibinfo
  {author} {\bibfnamefont {G.}~\bibnamefont {Tarjus}},\ }\href
  {https://doi.org/10.1103/PhysRevLett.87.055701} {\bibfield  {journal}
  {\bibinfo  {journal} {Phys. Rev. Lett.}\ }\textbf {\bibinfo {volume} {87}},\
  \bibinfo {pages} {055701} (\bibinfo {year} {2001})}\BibitemShut {NoStop}%
\bibitem [{\citenamefont {Kierlik}\ \emph {et~al.}(2002)\citenamefont
  {Kierlik}, \citenamefont {Monson}, \citenamefont {Rosinberg},\ and\
  \citenamefont {Tarjus}}]{kierlik2002}%
  \BibitemOpen
  \bibfield  {author} {\bibinfo {author} {\bibfnamefont {E.}~\bibnamefont
  {Kierlik}}, \bibinfo {author} {\bibfnamefont {P.~A.}\ \bibnamefont {Monson}},
  \bibinfo {author} {\bibfnamefont {M.~L.}\ \bibnamefont {Rosinberg}},\ and\
  \bibinfo {author} {\bibfnamefont {G.}~\bibnamefont {Tarjus}},\ }\href
  {https://doi.org/10.1088/0953-8984/14/40/319} {\bibfield  {journal} {\bibinfo
   {journal} {J. Phys. Condens. Matter}\ }\textbf {\bibinfo {volume} {14}},\
  \bibinfo {pages} {9295} (\bibinfo {year} {2002})}\BibitemShut {NoStop}%
\bibitem [{\citenamefont {Chen}\ \emph {et~al.}(2018)\citenamefont {Chen},
  \citenamefont {Coasne}, \citenamefont {Guyer}, \citenamefont {Derome},\ and\
  \citenamefont {Carmeliet}}]{chen2018}%
  \BibitemOpen
  \bibfield  {author} {\bibinfo {author} {\bibfnamefont {M.}~\bibnamefont
  {Chen}}, \bibinfo {author} {\bibfnamefont {B.}~\bibnamefont {Coasne}},
  \bibinfo {author} {\bibfnamefont {R.}~\bibnamefont {Guyer}}, \bibinfo
  {author} {\bibfnamefont {D.}~\bibnamefont {Derome}},\ and\ \bibinfo {author}
  {\bibfnamefont {J.}~\bibnamefont {Carmeliet}},\ }\href
  {https://doi.org/10.1038/s41467-018-05897-9} {\bibfield  {journal} {\bibinfo
  {journal} {Nat. Commun.}\ }\textbf {\bibinfo {volume} {9}},\ \bibinfo {pages}
  {3507} (\bibinfo {year} {2018})}\BibitemShut {NoStop}%
\bibitem [{\citenamefont {Lacks}(2012)}]{Lacks.2012}%
  \BibitemOpen
  \bibfield  {author} {\bibinfo {author} {\bibfnamefont {D.~J.}\ \bibnamefont
  {Lacks}},\ }\href {https://doi.org/10.1002/anie.201202896} {\bibfield
  {journal} {\bibinfo  {journal} {Angew. Chem., Int. Ed. Engl.}\ }\textbf
  {\bibinfo {volume} {51}},\ \bibinfo {pages} {6822} (\bibinfo {year}
  {2012})}\BibitemShut {NoStop}%
\end{thebibliography}%

\end{document}